# Phenotypic Plasticity, the Baldwin Effect, and the Speeding up of Evolution: the Computational Roots of an Illusion

Mauro Santos[1,*], Eörs Szathmáry [2,3,4,†], and José F. Fontanari [5,‡]

1. Departament de Genètica i de Microbiologia; Grup de Genòmica, Bioinformàtica i Biologia Evolutiva (GGBE); Universitat Autònoma de Barcelona; 08193 Bellaterra (Barcelona); Spain
2. Parmenides Center for the Conceptual Foundations of Science, Kirchplatz 1 Pullach, D-82049, Munich, Germany
3. Biological Insititute and Research Group in Evolutionary Ecology and Theoretical Biology, Eötvös University, Budapest, Hungary
4. Faculty of Biology, Ludwig Maximilians University Munich, Munich, Germany
5. Instituto de Física de São Carlos, Universidade de São Paulo, Caixa Postal 369, 13560-970 São Carlos SP, Brazil

[*] Corresponding author; e-mail: mauro.santos@uab.es
[†] E-mail: szathmary.eors@gmail.com
[‡] E-mail: fontanar@ifsc.usp.br

**Abstract**

An increasing number of dissident voices claim that the standard neo-Darwinian view of genes as 'leaders' and phenotypes as 'followers' during the process of adaptive evolution should be turned on its head. This idea is older than the rediscovery of Mendel's laws of inheritance and has been given several names before its final 'Baldwin effect' label. A condition for this effect is that environmentally induced variation such as phenotypic plasticity or learning is crucial for the initial establishment of a population. This gives the necessary time for natural selection to act on genetic variation and the adaptive trait can be eventually encoded in the genotype. An influential paper published in the late 1980s showed the Baldwin effect to happen in computer simulations, and claimed that it was crucial to solve a difficult adaptive task. This generated much excitement among scholars in various disciplines that regard neo-Darwinian accounts to explain the evolutionary emergence of high-order phenotypic traits such as consciousness or language almost hopeless. Here, we use analytical and computational approaches to show that a standard population genetics treatment can easily crack what the scientific community has granted as an unsolvable adaptive problem without learning. The Baldwin effect is once again in need of convincing theoretical foundations.

**1.    Introduction**

What role does the Baldwin effect play in evolution? By Baldwin effect −term coined by Simpson [1]− we refer to a turn-of-the-twentieth-century idea [2,3,4] cogently described by Maynard Smith [5, p. 761] as follows: "If individuals vary genetically in their capacity to learn, or to adapt developmentally, then those most able to adapt will leave most descendants, and the genes responsible will increase in frequency. In a fixed environment, when the best thing to learn remains constant, this can lead to the genetic determination of a character that, in earlier generations, had to be acquired afresh in each generation". The Baldwin effect involves two transitions [6,7]: the first has to do with the evolutionary value of phenotypic plasticity, or some particular form of plasticity such as learning; the second with the 'genetic accommodation' (i.e., evolution in response to both genetically based and environmentally induced novel traits [8,9,10]) of the learned trait. We use genetic accommodation instead of the more familiar term 'genetic assimilation' coined by Waddington [11] because this last term should not be equated to the Baldwin effect [10, see also 9, pp. 153-154].

Some towering figures in the Modern Synthesis −expression borrowed from the title of Julian Huxley's [12] book− were either indulgent with the theoretical plausibility of the Baldwin effect [1] or utterly hostile towards it, recommending to discard this concept altogether [13,14]. This advice is followed suit by several influential textbooks in evolutionary biology [15,16,17] that do not even mention Baldwin at all. However, although at present there appears to be no clear empirical evidence for Baldwin effects, several authors have called for a radical revision of the consensus view and argued that much evolution involves genetic accommodation [9,18,19,20; but see 21]. The current tension among evolutionary biologists [22] is unmatched by evolutionary computationalists [6,23,24] and scholars in others disciplines (typically evolutionary psychologists and cognitive scientists [25]), who invoke the Baldwin effect as a major evolutionary force that could have led to the emergence of mind [26,27,28] and to modern language [29,30,31,32]. As Yamauchi [33, p. 3] put it, "the Baldwin effect is particularly appealing because … It may provide a natural Darwinian account for language evolution: It is an especially popular idea among linguists that language evolution is somehow saltational. This leads them to conclude *neo*-Darwinian theories are 'incompetent' for accounting for language evolution" (our addition in italics). (Neo-Darwinism is used here to describe the Modern Synthesis version of Darwinism.)

Much of the recent 'excitement' with the Baldwin effect stems from a seminal paper published by computer scientists Geoffrey Hinton and Steven Nowlan in the late 1980s [34], which has been cited 1,081 times (Google Scholar) to date. They developed a computational model combining a genetic algorithm with learning by trial and error in a sexual population of chromosomes (the 'organisms') that were initially segregating at $L = 20$ loci with three alleles each: 1, 0, and ?. This chromosome determines the connectivity of a neural network: allele 1 at a given locus indicates that a particular connection exists whereas allele 0 at that locus indicates that it does not. The question marks are plastic alleles that allow the organism to set (or not) the connection at the end of a learning period. The neural network has only one correct configuration of connections and the task the organisms had to solve was to find this configuration out of the $2^L \approx 10^6$ possible configurations. We can assume without loss of generality that the right answer is the chromosome with all alleles1; i.e., a fully connected neural network. The catch is that any other configuration provides no information whatsoever about where the correct answer might be. In such problems, there is no better way to search than by exhaustively sampling the entire combinatorial space; a situation termed a 'needle-in-the-haystack' problem. In other words, there is no efficient algorithm that can find the fitness maximum unless we introduce some 'trick'; namely, to somehow smooth the spiked fitness landscape through phenotypic plasticity [35]. Hinton and Nowlan [34] assumed that each organism could try up to a maximum of $G = 1,000$ random guesses for the settings of the ? states; these alleles define the 'plastic genome'. The organisms were also given the ability to recognize whether they have found the correct settings after $g < G$ learning trials and, in such a case, stop guessing (see below for details). Therefore, those organisms that were relatively fast at learning the correct configuration of alleles enjoyed a fitness advantage and produced more offspring. In the long run (well before 50 generations in the simulation performed by Hinton and Nowlan [34]), natural selection redesigned the genotypes in the population and the correct alleles 1 increased in frequency. Nonetheless, they did not take over and undecided alleles ? remained segregating at relatively high frequency because in the end organisms were able to learn quickly and, therefore, there was not much selective pressure to fix the 'innately correct' fitter alleles.

The scenario in Hinton and Nowlan [34] showed (i) that the Baldwin effect can be observed *in silico*, and (ii) that learning can dramatically accelerate adaptive evolution in a flat fitness landscape with a single isolated peak; what Ancel [36, p. 307] characterized as the "Baldwin expediting effect". Maynard Smith [5, p. 762] explained this effect by making a simple contrast with a population where organisms do not learn: "In a sexual population of 1,000 with initial allele frequencies of 0.5, a fit individual would arise about once in 1,000 generations … Mating would disrupt the optimum genotype, however, and its offspring would have lost the adaptation. In effect, a sexual population would never evolve the correct settings… (or does so excessively slowly)". Actually, "the problem was never solved by an evolutionary search without learning" [34, p. 497]. Conversely, Maynard Smith [5] claimed that in the absence of learning a large asexual population would include optimal individuals and the correct settings would soon be established by selection.

The first claim about non-learning sexual organisms has been taken for granted, whereas the second claim concerning asexual organisms was analytically investigated by Fontanari and Meir [37] to answer the question: how soon is 'soon'? Using their recursion equation (3.1) to analyze the evolution of correct alleles, the answer is that it would take more than 3,000 generations for the population to evolve the correct settings with initial allele frequencies 0.5 and no mutation. Therefore, the conclusion seems to be fairly clear: in the single-peaked fitness landscape assumed by Hinton and Nowlan [34] learning has a drastic effect on evolution.

Here, we show that this conclusion is generally incorrect and requires careful considerations. The heart of the problem was also pointed out by Maynard Smith [5] and relates to the strong positive epistasis in Hinton and Nowlan's [34] scenario. This epistasis generates, in turn, strong positive associations between the correct alleles in the non-learning organisms that can greatly accelerate evolution (Appendix A). The former solution of more than 3,000 generations for the asexual population to evolve the correct settings is likely to be a gross overestimate as Fontanari and Meir [37] ignored the generation of linkage disequilibrium due to directional selection. The remainder of the paper is organized as follows. First, we discuss Hinton and Nowlan [34] model in more detail as it will make the reason for our skepticism about what they have really demonstrated very clear. Second, we derive the exact recursion equations for the asexual case and show that evolution is indeed quite fast in this case. Third, challenging the conventional wisdom we show that a finite population of sexual organisms that do not learn does evolve the correct settings, and estimate the probability of fixation and mean time to fixation of the correct genotype as a function of population size $N$ and chromosome length $L$. Finally, we summarize our results and point out the misconceptions generated by the computer simulations in Hinton and Nowlan [34].

## 2. The simulation by Hinton and Nowlan

Hinton and Nowlan's [34] basic idea was to show that a haploid sexual population of organisms with plasticity (learning ability) will evolve towards an optimal phenotype in fewer generations than a population of organisms that do not learn. They assumed that the $L = 20$ loci code for neural connections and alleles 1 specify innately correct connections, alleles 0 innately incorrect connections, and alleles ? guessable (plastic) connections containing a switch that can be on (right) or off (wrong). Learning consists of giving each individual up to a maximum of $G = 1,000$ random combinations of switch settings (with equal probability for on and off) on every trial. Those individuals that have a 0 allele at any locus will never produce the right connectivity of the neural network. On the other hand, if the combination of switch settings and the genetically specified connections produce the good net (i.e., a fully connected neural network) after $g < G$ trials the individual stops guessing.

Hinton and Nowlan's [34] evolutionary algorithm performs two operations. First, each organism $(i = 1, 2, ..., N)$ is evaluated according to fitness, which determines the mating probability and offspring production according to the following fitness function

$$w_i = 1 + \frac{(L-1)\eta_g}{G}, \qquad (1)$$

where $\eta_g = (G - g)$ is the number of trials remaining after the correct configuration of connections has been found. This fitness function is central to Hinton and Nowlan [34] argument. It indirectly smoothes out the landscape and the organisms nearby the attraction zone of the peak (i.e., those chromosomes with ones and question marks that can be correctly set after $g < G$ trials) enjoy increased fitness. The basal fitness is $w_i = 1$ if the organism never gets the right answer after $G$ trials, and has a maximum value of $w_i = L$ for an organism that already has all its connections innately specified.

Second, the crossover operation picks one point $(1 \leq m \leq L-1)$ at random from each of parents' chromosomes to form one offspring chromosome by taking all alleles from the first parent up to the crossover point, and all alleles from the second parent beyond the crossover point. Although not explicitly stated in Hinton and Nowlan [34], taking $m \leq L-1$ as the upper bound guarantees that the offspring will always be a recombinant string. None of the learning is passed on to children, which inherit the same allelic configuration their parents had at the different loci. Hinton and Nowlan [34] simulation is replicated in Fig. 1.

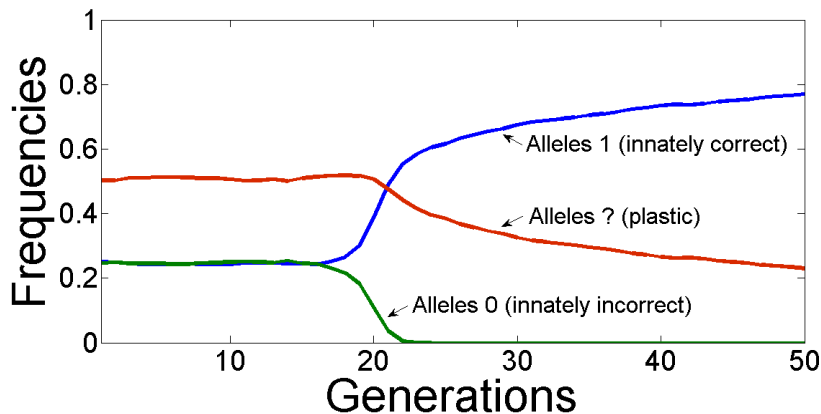

**Fig. 1. Replication of Hinton and Nowlan's original simulation.** The population is able to quickly learn the solution as innately incorrect alleles 0 are eliminated from the population, although the frequency of plastic alleles ? remains relatively high.

## 3. Evolution without learning in an asexual population: equations for the deterministic limit

We will consider the case where learning is absent in order to show that the neglect of linkage disequilibrium [37] greatly overestimates the number of generations required for the correct genotype to become preponderant in the population. In fact, in an infinite asexual population the prevalence of the correct genotype takes place so rapidly that the effect of learning (if any) is not significant.

In this case the genotypes are binary strings of size $L$ because without learning there is no difference between alleles $0$ and $?$. There are $2^L$ different such strings and we denote their frequencies in the infinite population at generation $t$ by $Y_\alpha(t)$, with $\alpha = 1, \ldots, 2^L$. Without loss of generality, we will assume that the correct genotype, i.e. the string $(1, 1, \ldots, 1)$, corresponds to $\alpha = 1$. Recalling that the fitness $w_\alpha$ of all genotypes but the correct one is set to the baseline value $w_{\alpha \neq 1} = 1$, and that the fitness of the correct genotype is set to $w_1 = L$, we obtain

$$Y_1(t) \approx 1 - \frac{1 - Y_1(0)}{[Y_1(0)]^L} \exp\left[-(L-1)t/L\right], \quad (2)$$

in the regime $t \gg 1$ for which $Y_1(t) \approx 1$ (see Appendix B). This expression allows us to estimate the number of generations needed for $Y_1(t)$ to attain some arbitrary value close to one. Solving Eq. $(2)$ for $t$ yields

$$t \approx -\frac{1}{L-1} \ln\left\{Y_1(0)\left[\frac{1-Y_1(t)}{1-Y_1(0)}\right]^L\right\}, \quad (3)$$

from which we see that $t$ increases with the logarithm of $Y_1(0)$ rather than with a negative power of $Y_1(0)$ as in the case where linkage disequilibrium is neglected [37].

At this stage, it is instructive for comparative purposes with previous work [37] to derive the corresponding equations for the evolution of an infinite asexual population ignoring the generation of linkage disequilibrium. This is important because the paper by Fontanari and Meir [37] has routinely been cited as giving additional support for the benefits of learning, without realizing that their analytical treatment was inappropriate for an asexual population. Their derivation begins with the expression for the frequency of alleles 1 at generation $t+1$, which we denote by $p(t+1)$, given that one knows the genotype frequencies at generation $t$; i.e.,

$$p(t+1) = \frac{1}{L} \frac{\sum_\alpha P_\alpha w_\alpha Y_\alpha(t)}{\sum_\alpha w_\alpha Y_\alpha(t)}, \qquad (4)$$

were $P_\alpha$ stands for the number of alleles 1 in string $\alpha$. Ignoring linkage disequilibrium transforms this equation into an autonomous recursion equation for $p(t)$ by assuming that $Y_\alpha(t) = [p(t)]^{P_\alpha}[1-p(t)]^{L-P_\alpha}$ for all $\alpha$ and $t$. This means that the abundance of a genotype depends only on the number of alleles 1 and not on the specific location of those alleles in the string. In addition, those abundances are completely determined by the global frequency of alleles 1 in the population according to the previous expression. It is as if all genotypes were disassembled and then reassembled again at random following a procedure akin to Wilson's [38]; that is, Fontanari and Meir [37] wrongly assumed $L-1$ recombination points. In fact, the reason why ignoring linkage disequilibrium is an inappropriate and uncontrolled approximation in this context, as well as its connection to group selection, was already pointed out in Alves et al. [39, section V]. This neglect gives the following incorrect recursion equation [37; equation (3.1) ignoring mutation] after inserting $Y_\alpha(t)$ into Eq. (4)

$$p(t+1) = \frac{p(t) + (L-1)[p(t)]^L}{1 + (L-1)[p(t)]^L}. \qquad (5)$$

For $L=20$ and $p(0) = 0.25$ (i.e., $Y_1(0) = 1/2^{40}$), the numerical solution of Eq. (5) shows that it takes about $t = 5.2 \times 10^8$ generations to reach the regime where $1 - p(t) = 10^{-6}$. This was the reason why Fontanari and Meir [37] concluded that learning has a drastic effect on evolution in Hinton and Nowlan's [34] scenario. However, using the correct Eq. (3) we find that only $t = 15$ generations are sufficient to reach that regime.

In the same vein than Hinton and Nowlan [34] we did not include mutation in the foregoing treatment, but it is easy to derive exact recursion equations for the non-learning asexual case assuming mutation (Appendix C). In summary, the preponderance of the correct genotype in an infinite asexual population without learning takes place extremely fast so the effect of learning, if any, is not significant in this case.

## 4. Evolution without learning in a sexual population

Having shown that Maynard Smith's [5, p.762] remark that a non-learning asexual population of "many millions would breed true, and the correct settings would soon be established by selection" ('soon' is indeed very soon) is essentially correct, what of a sexual population? In this context, it is interesting to recall what Hinton and Nowlan [34, p. 497] wrote (our emphasis in italics): "The same problem was never solved by an evolutionary search without learning. This was not a surprising result; the problem was selected to be extremely difficult for an evolutionary search, which relies on the exploitation of small co-adapted sets of alleles to provide a better than random search of the space… To preserve the co-adaptation from generation to generation it is necessary for each good genotype, on average, to give rise to at least one good descendant in the next generation. If the dispersal of complex co-adaptations due to mating causes each good genotype to have less than one expected good descendant in the next generation, the co-adaptation will not spread, even if it is discovered many times. *In our example, the expected number of good immediate descendants of a good genotype is below 1 without learning and above 1 with learning*." This conclusion was also repeated by Maynard Smith [5, p. 762]: "In a sexual population of 1,000 with initial allele frequencies of 0.5, a fit individual would arise about once in 1,000 generations … Mating would disrupt the optimum genotype, however, and its offspring would have lost the adaptation. In effect, a sexual population would never evolve the correct settings".

Here, we show that this conclusion is incorrect. Some hints why this is mistaken can be gained from Eq. $(5)$, which basically assumes that all genotypes are disassembled and then reassembled again at random following a procedure akin to Wilson's [38] trait group selection framework; namely, Eq. $(5)$ assumes $L-1$ recombination points as already stated above. This wrong assumption notwithstanding, an eventual fixation of the correct string is attained [37].

A more straightforward demonstration that a non-learning sexual population can find the solution to the 'needle-in-the-haystack' problem is obtained by showing that the expected number of good immediate descendants of a good genotype is actually above 1, in stark contrast to Hinton and Nowlan's claim quoted before. In fact, assuming that the correct all 1s string is present in the population at some generation, we can easily calculate the distribution of the number of good offspring it generates by mating with a random string following the crossover operation in Hinton and Nowlan [34]. As indicated above, the single offspring of each mating is generated by randomly choosing a crossover point and taking all alleles from the first parent up to the crossover point, and from the second parent beyond the crossover point. Let us take the all 1s string as the first parent and pick another string at random from the $2^L$ possible strings as the second parent. The probability that the resulting offspring is an all 1s string is simply

$$\rho = \frac{1}{L-1}\sum_{i=1}^{L-1}\frac{1}{2^i} = \frac{1}{L-1}\left(1-\frac{1}{2^{L-1}}\right) \approx \frac{1}{L-1}, \tag{6}$$

since all crossover points are equiprobable and the second parent must be all 1s after the crossover point. Therefore, the mean number of good offspring produced in $L$ mates (the fitness of the correct genotype is $w_1 = L$) is $L\rho = L/(L-1) > 1$ and so the "expected number of good immediate descendants of a good genotype" is not below one as claimed [34; see also 5] but, quite the opposite, the good genotype is expected to increase exponentially once it has appeared in the population. This is shown in Fig. 2 using the same parameter values (population size $N = 1,000$, chromosome length $L = 20$) than Hinton and Nowlan [34] with allele frequencies 0.5, where the correct genotype went to fixation in 14 out of 100 independent runs (14%) in less than 150 generations. Obviously, the former conclusion that a non-learning sexual population would never evolve the correct settings is just plain wrong.

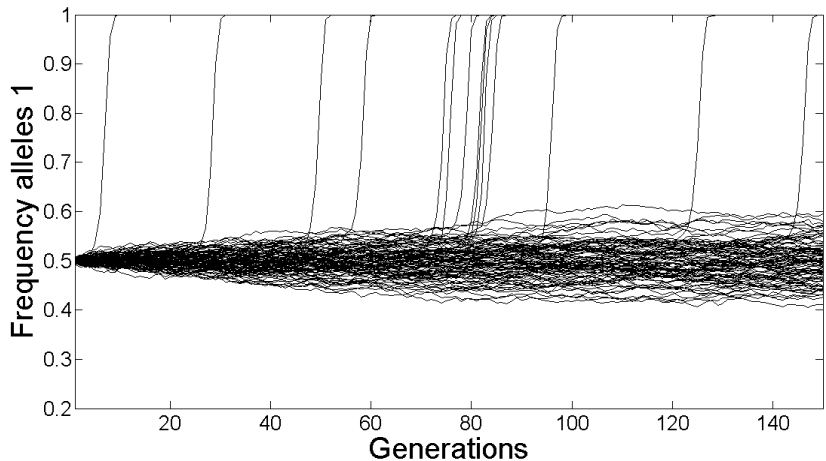

**Fig. 2. Evolutionary search without learning in a sexual population.** The simulations follow Hinton and Nowlan [34] scheme with the exception that each organism is a binary string. Initial frequencies were 0.5,

population size $N = 1,000$ and chromosome length $L = 20$. A total of 100 independent runs were followed for 150 generations, and fixation of the correct genotype was observed in 14 runs (14%). This fixation is conditional on the first appearance of the correct genotype in the population; once it appears, its frequency increases exponentially. With the parameters values used, the probability of occurrence of the correct genotype at the initial generation is equal to $1/2^{20} \times 1,000 = 9.5367 \times 10^{-4}$. Therefore, when $N/2^L \ll 1$ the probability of fixation of the correct genotype is mostly dependent on the balance between the mean time to its first appearance by recombination and the mean time to fixation by genetic drift of the incorrect allele 0 at any single locus; an event that prevents the fixation of the correct genotype (mutation was ignored in these simulations).

As an application of Eq. $(6)$, we can derive the mean number of good offspring in a two-generation dynamics of a population of size $N$. Assume the population at the first generation is composed of a single correct genotype (fitness $L$) plus $N-1$ random genotypes (fitness 1). Since two distinct parents are chosen in each mating, the probability that the good genotype is chosen is

$$\frac{L}{L+N-1} + \frac{N-1}{L+N-1} \times \frac{L}{L+N-2}, \qquad (7)$$

where the first term is the probability that the good genotype is the first chosen mate and the second term is the probability that it is the second mate. As a generation comprises $N$ such mates and the probability of resulting a good offspring is $\rho$, we obtain that the mean number of good offspring in the second generation is

$$\mu = \frac{NL}{(L+N-1)(L-1)}\left[1 + \frac{N-1}{L+N-2}\right]. \qquad (8)$$

Since we have assumed that good offspring are produced only by mates involving the good genotypes and not by the recombination of random strings, Eq. $(8)$ yields a lower bound for $\mu$ but it fits the simulation data very well in the regime $N \ll 2^L$ (data not shown). Most interestingly, for large populations $N \gg L$ we find that $\mu$ tends to 2 regardless the value of $L$, which explains the very fast growth of the good genotype once it appears in the population (Fig. 2).

Our findings open a number of questions. What is the probability of fixation of the correct string as a function of $N$ and the search space $2^L$? What is the expected number of generations for the correct genotype to reach fixation? What is the scaling factor between $N$ and $L$ for successful fixation? The reason why these questions are important is the following. As discussed by Belew [40], Hinton and Nowlan [34] picked the parameters in their simulations very carefully as they assumed a population size of $N = 1{,}000$ organisms, and allele frequencies of 0.25 for zeros, 0.25 for ones, and 0.50 for question marks. With $L = 20$, on average half of the alleles will be ? and there are $2^{10} = 1{,}024$ combinations to try. Therefore, it is no surprise that given 1,000 organisms and up to $G = 1{,}000$ learning trials per organism the correct settings for the ? connections were easily found. If, for instance, we keep $G$ constant and let the chromosome length $L$ to increase from 20 to 30, the time taken for the population to solve the task in Hinton and Nowlan [34] scenario increases exponentially [37]: with $G = 1{,}000$ and $L = 30$ it would take more than $10^3$ generations before an arbitrary large population of learning organisms can find the solution (results not shown). Therefore, both $G$ and $L$ need to be cautiously chosen (i.e., $G/2^{L/2} \approx 1$) to sustain the claim that learning speeds up evolution in a single-peaked fitness landscape. Nonetheless, to the extent that the scientific community accepted that problem was unsolvable in any reasonable fashion without learning [e.g., 5, 26 pp.77-80, 28 p. 178, 41,42], this could be considered as a relatively 'minor' detail in Hinton and Nowlan [34] simulations. After all, they provided a proof of concept and Baldwin effects seemed to be essential to solve their difficult adaptive task, which has generated a growing scientific literature ever since. However, the demonstration that a population of non-learning sexual organisms can also find the needle in the haystack converts what appeared to be a qualitative issue into a quantitative problem in its own right (Fig. 3).

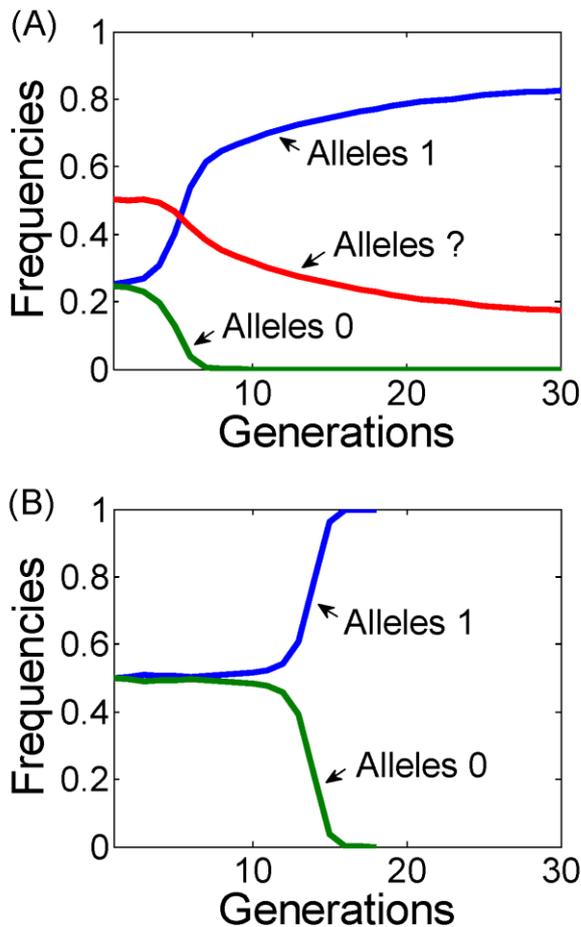

Fig. 3. Evolutionary search with (A) and without (B) learning in a sexual population. (A) the simulation uses Hinton and Nowlan [34] algorithm with allele frequencies of 0.25 for zeros, 0.25 for ones, and 0.50 for question marks; but different parameter values for population size ($N = 2,500$), chromosome length ($L = 16$), and maximum number of learning trials per organism ($G = 250$). With $L = 16$ there are $2^8 = 256$ combinations to try as an average for the settings of ? alleles; that is, we kept the same relationship $G/2^{L/2} = 0.977$ than Hinton and Nowlan's original simulation. (B) the simulation follows Hinton and Nowlan [34] scheme with the exception that each organism is a binary string. Initial frequencies were 0.5, population size $N = 2,500$ and chromosome length $L = 16$. The point here is that in this scenario the claim that learning allows organisms to evolve much faster than their non-learning counterparts does not seem to be fully justified.

### 4.1 Probability of fixation

We used computer simulations to estimate the probability of fixation of the correct genotype (denoted as $P_1$). The simulations followed Hinton and Nowlan [34] scheme. In particular, the population consists of $N$ binary strings of length $L$ and update is parallel; i.e., generations do not overlap. To create the next generation from the current one, we perform $N$ matings. The two parents of a mating are different individuals that are chosen at random from the current generation with probability proportional to fitness ($w_1 = L$ is the fitness of the correct genotype and $w_{\alpha \neq 1} = 1$ is the fitness of all genotypes but the correct one). The single offspring of each mating is generated after applying the one point crossover operation (see above). The initial population is generated randomly by choosing the $L$ digits of each string as 0 or 1 with equal probability.

For different string lengths, the probability of fixation $P_1$ as a function of $N$ is plotted in Fig. 4A, and as a function of the rescaled variable $\xi = N^{1.9}/2^L$ in Fig. 4B.

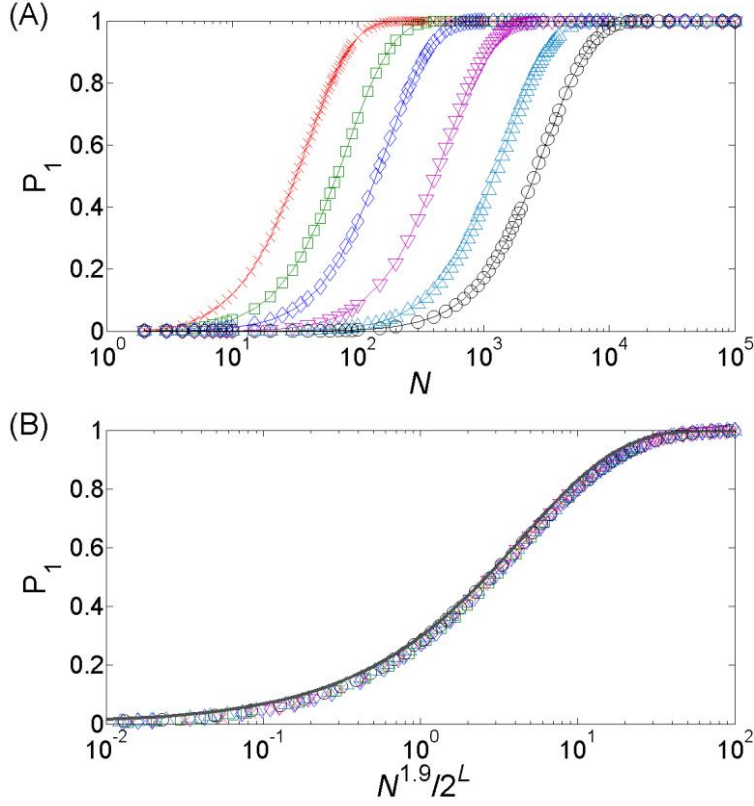

**Fig. 4. Probability of fixation $(P_1)$ of the correct genotype in a non-learning sexual population as a function of population size $N$.** (A) plots from left to right chromosome lengths $L = 8$ (red, times signs), 10 (green, squares), 12 (blue, diamonds), 15 (magenta, inverted triangles), 18 (cyan, triangles) and 20 (black, circles). Each symbol represents the fraction of the simulations in which we observed the fixation of the correct genotype and the lines are guides to the eyes. For $N = 1{,}000$ and $L = 20$ we find $P_1 = 0.172$. (B) is the same but the probability of fixation is plotted against the rescaled variable $\xi = N^{1.9}/2^L$. The solid grey line is the fitting function $f(\xi) = 1 - \exp(-0.353 \times \xi^{0.7})$.

The probability $P_1$ is estimated as the fraction of simulations in which we observed the fixation of the correct genotype. The number of simulations varied from $10^7$ to $10^4$ so as to guarantee that a statistically significant number of correct fixations have occurred. The variable $\xi$ reveals the way $N$ must scale with $L$ in order to maintain $P_1$ invariant, resulting thus in the collapse of the data of Fig. 4A into a single universal scaling function, which seems to be well approximated by the fitting function $f(\xi) = 1 - \exp(-0.353 \times \xi^{0.7})$ as shown in Fig. 4B. We note that for a *finite asexual* population the corresponding scaling is given by $\xi = N/2^L$ (data not shown). Thus, with $L = 20$ fixation $(P_1 \geq 0.99)$ of the correct genotype in the sexual case is almost guaranteed when $N \geq 10{,}170$; that is, with a population size well below the size of the solution space $\left(N/2^L = 0.0097\right)$. The reason for this can be understood from the following heuristic argument. Assume for simplicity that once the all 1s string appears in the population its eventual fixation will occur because its frequency will increase exponentially (see above). With $N < 2^L$, the correct genotype will be present at the initial population with probability $N/2^L$, which is quite low for $N = 10,170$ and $L = 20$. It will likely arise through recombination with the same probability $N/2^L$ at each generation (we ignore the fact that allele frequencies will drift apart from their initial frequency of 0.5), which means that the probability of no occurrence of the correct genotype decreases with the number of generations as $\left(1 - N/2^L\right)^t$. With a large enough population size genetic drift will not be very important and the correct genotype will eventually appear and spread to fixation. Maynard Smith [5] previous argument needs to be rewritten as follows: In a sexual population of 1,000 with initial allele frequencies of 0.5, a fit individual would arise about once in 1,000 generations. Once it appears, it will reach fixation in few generations (actually, the probability of fixation in this case is $P_1 = 0.172$; Fig. 4A).

### 4.2 Mean time to fixation

The previous computer simulations also allowed estimating the (conditional) mean time to fixation of the correct genotype (denoted as $T_1$). For different string lengths, $T_1$ is plotted as a function of the number of strings $N$ in Fig. 5A, and as a function of the ratio between $N$ and the size of the solution space $N/2^L$ in Fig. 5B. For $L > 20$ we find that the height of the $T_1$ peak increases as $2^{L/2}$ which, most interestingly, coincides with the form the number of learning trials $G$ scales with the string length $L$. In particular, for $L = 20$ we find (mean ± SD) $T_1 = 174 \pm 142$ generations with $N = 10^3$, and $T_1 = 28 \pm 14$ generations with $N = 10^5$. In other words, when $N/2^L \approx 0.1$ the mean time to fixation of the correct genotype is similar to Hinton and Nowlan [34] situation with learning (Fig. 1). In general, for fixed $L$ and very large $N$ we have that $T_1$ grows with $\log N$ and is practically insensitive to $L$. This is in sharp contrast with what happens with learning because both $L$ and $G$ have to increase simultaneously $\left(G/2^{L/2} \approx 1\right)$, otherwise the time taken for the population to solve the task in Hinton and Nowlan [34] scenario increases exponentially [37] as pointed out above.

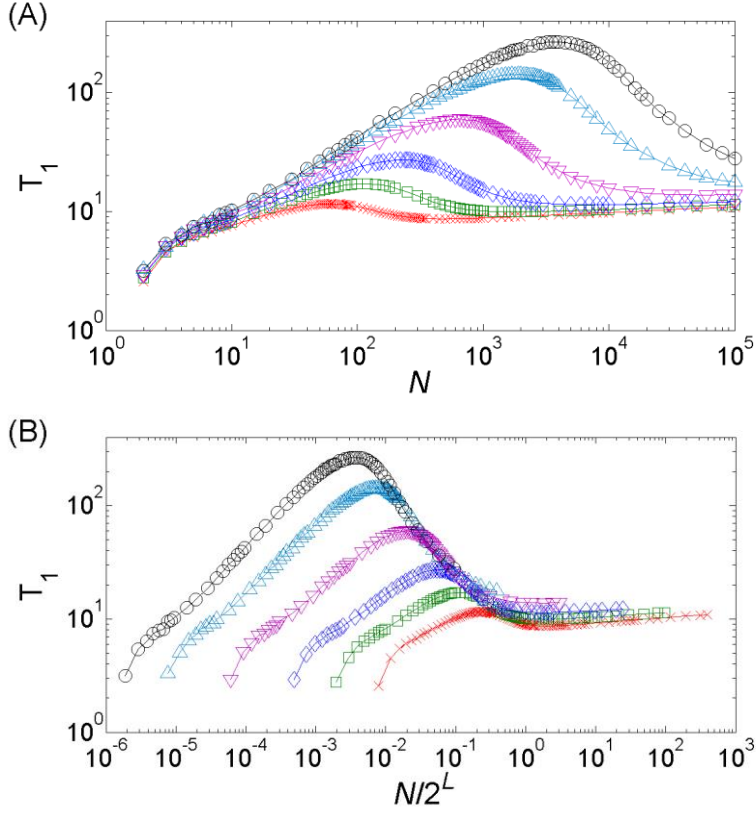

Fig. 5. Conditional mean time to fixation $(T_1)$ of the correct genotype in a non-learning sexual population as a function of population size $N$. (A) plots from bottom to top chromosome lengths $L=8$ (red, times signs), 10 (green, squares), 12 (blue, diamonds), 15 (magenta, inverted triangles), 18 (cyan, triangles) and 20 (black, circles). For $L=20$ we find (mean ± SD) $T_1 = 174 \pm 142$ generations with $N = 10^3$, and $T_1 = 28 \pm 14$ generations with $N = 10^5$. (B) is the same but the mean time to fixation is plotted against the ratio between the population size and the size of the solution space $(N/2^L)$. The lines are guides to the eyes.

To sum up, in the single-peaked fitness landscape learning speeds up evolution whenever the ratio between the maximum number of allowed guesses per organism $G$ and the size of the 'guessing space' $2^{L/2}$ is on the same order, i.e. $G/2^{L/2} \approx 1$, and the ratio between population size $N$ and the size of the solution space $2^L$ in non-learning organisms is $N/2^L < 0.1$. That is, with $G = 1,000$, $L = 20$ and 1,000 organisms as in Hinton and Nowlan [34] the $10^6 \approx 2^L$ queries to the fitness function per generation accomplish the same result than sampling $10^5 \approx 2^L/10$ genotypes per generation in the non-learning situation (one extra genotype is worth 10 queries). Decrease the ratio $G/2^{L/2}$ below 1 and keep the ratio $N/2^L$ around 0.1 and the result is that non-learning sexual organisms will start doing better than their learning equivalents.

## 4.3 Exploring other models of recombination

The successful fixation of the good genotype once it appears in the population is critically dependent on the expected number $L\rho$ of good offspring after recombination, where $\rho$ is the probability that any single offspring is an all 1s string. As we have previously shown, $L\rho > 1$ in the one point recombination scenario assumed by Hinton and Nowlan [34]. Our question now is to what extent this result is robust to a more general recombination model.

Here we follow the stochastic multilocus method of Fraser and Burnell [43] and model recombination as a random walk along the length of the two parental chromosomes, changing from one to the other within the constraint of the probability of such a change; namely, the recombination rate among two adjacent loci. Under the assumption that the probability of recombination between any two loci is constant and equal to $r$, the probability that a mate between a $L$ loci good genotype and a random genotype produces a good offspring is:

$$\rho = (1-r)\rho_1 + r\rho_2, \qquad (9)$$

where $\rho_1$ is the probability that the offspring is the good genotype when the random walk starts from the all 1s parent, and $\rho_2$ is the same probability when the random walk starts from the other random parent. Here

$$\rho_1 = (1-r)^{L-1} + \sum_{k=1}^{L-1} r^k (1-r)^{L-1-k} \sum_{i=0}^{L-M} \binom{O-1+i}{i}\binom{L-i-O-1}{L-M-i} 2^{i+O-L}, \qquad (10)$$

and

$$\rho_2 = (1-r)^{L-1} 2^{-L} + \sum_{k=1}^{L-1} r^k (1-r)^{L-1-k} \sum_{i=0}^{L-M} \binom{E-1+i}{i}\binom{L-i-E-1}{L-M-i} 2^{i+E-L}, \qquad (11)$$

where $M = k+1$, $O$ is the number of odd numbers in the sequence 1, 2, 3,..., $M$, i.e., $O = M/2$ if $M$ is even and $O = (M+1)/2$ if $M$ is odd; and $E$ is the number of even numbers in the same sequence, i.e., $E = M/2$ if $M$ is even and $E = (M-1)/2$ if $M$ is odd. As before, the mean number of good offspring genotypes is $L\rho$; the product between the fitness of the good parental genotype and $\rho$.

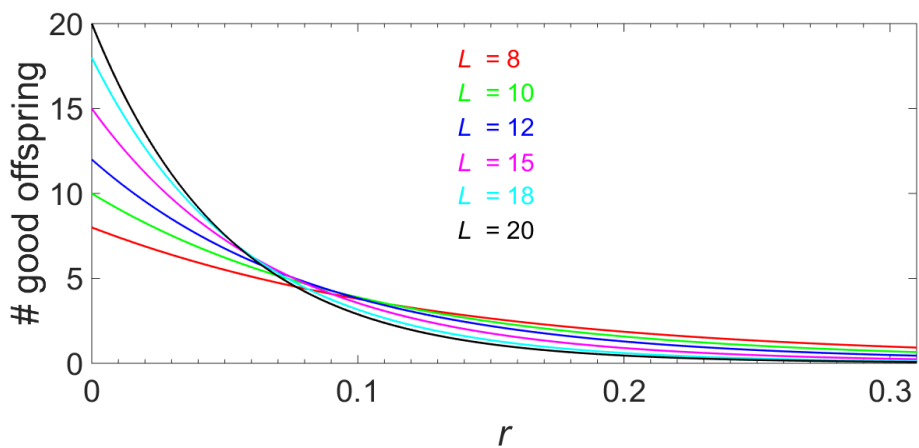

**Fig. 6. Expected number of good offspring when an all 1s parental chromosome mates with $L$ random chromosomes.** The curves plot $L\rho$, where $\rho$ is the probability that any single offspring is an all 1s string given by eq. (9) following the stochastic multilocus recombination method of Fraser and Burnell [43].

Eqs. (10) and (11) allow us to set an upper bound for $r$ such that $L\rho > 1$ (Fig. 6). These results agree perfectly with simulation results (not shown). Thus, with $L = 20$ the upper bound is around $r = 0.156$, which is a large recombination rate between adjacent loci. For instance, a recombination frequency of only $r = 10^{-3}$ in our analysis would correspond to a chromosome with 1,000 loci and a map length of approximately 100 centimorgans; about the map length of each of the two major chromosomes of *Drosophila melanogaster* [44], which together contain about 80% of the species' genome. Therefore, the conclusion that once the good string appears recombination would do little harm to its spread by selection seems to be robust. This conclusion was checked by performing 100 independent simulation runs as in Fig. 2 but now using the stochastic multilocus recombination method of Fraser and Burnell [43] with $r = 0.15$. The correct genotype went to fixation in 12 runs (12%) in less than 150 generations (results not shown).

## 5. Conclusions

As pointed out by Dennet [45], many scholars including himself thought that Hinton and Nowlan [34] and Maynard Smith [5] had shown clearly and succinctly how and why the Baldwin effect worked, to the extent that in his bestseller book "Consciousness Explained" Dennet [46, p. 186] wrote that "thanks to the Baldwin effect, species can be said to pretest the efficacy of particular different designs by phenotypic (individual) exploration of the space of nearby possibilities". We grant this sort of claims as misleading for two reasons.

First, it takes for granted the happening of an effect that still waits for convincing empirical support after more than 100 years since its original inception. Second, it disregards the countless times proved effectiveness of standard neo-Darwinian selection to evolve complex biological traits [16,47] and contribute to some of the major evolutionary transitions [48].

Yeh and Price [49] speculated that colonization and establishment of a new population of dark-eyed juncos in coastal California was facilitated by a plastic response in breeding season, and claimed that their results provide the first quantitative evidence of Baldwin's proposition that plasticity aids individuals to deal with novel situations. Baldwin [2] certainly deserves the merit of "setting out a nascent theory of the evolution of phenotypic plasticity" [50, p. iii]; however, it should be remembered that the Baldwin effect involves two transitions: the evolutionary value of phenotypic plasticity and the genetic accommodation of the induced trait. Because Yeh and Price [49] did not prove that breeding season −a highly plastic trait [51]− was genetically accommodated, we do not think their paper can be cited as a clear example of Baldwin effects [10]. Simpson's [1] careful scrutiny made on the plausibility of Baldwin effects also applies here.

Our aim here was not to dismiss Hinton and Nowlan's [34] seminal contribution persuasively demonstrating the feasibility of the Baldwin effect. Their short paper offers *pima facie* evidence of Haldane's conviction that "if you are faced by a difficulty or a controversy in science, an ounce of algebra is worth a ton of verbal argument" [52, p. 239], which is undoubtedly the reason behind the huge positive influence of their work. Hinton and Nowlan [34] showed how the Baldwin effect could happen, but our present results show that there is no need for Baldwin effects to happen − at least in the proposed scenario.

It is critical to clarify what we have and have not shown here. We have shown that finding a needle-in-the-haystack without learning is a trivial enterprise for a large population of asexual organisms, as well as for a sexual population depending on the scaling factor between population size and the search space. Therefore, Hinton and Nowlan's [34] 'genes as followers' scenario [9] for the Baldwin effect could also be reframed into the traditional (Modern Synthesis) perspective that genes are 'leaders' and phenotypes are 'followers' during the process of adaptive evolution.

We have not shown that Baldwin effects are unlikely to happen in nature. Although we are not completely hostile but share Simpson's [1] skepticism about the concept, to prove or disprove Baldwin effects is ultimately an empirical question. It should also be clarified that our doubts on the actual relevance of Baldwin effects should not be taken as a criticism to the role of behavior in evolution [53,54], as it is already obvious by contrasting Mayr's [13] hostility towards Baldwin effects and his vindication of behavior as an important pacemaker or driver of evolutionary change [55]. Actually, one of us has conjectured that behavioral thermoregulation has been responsible for the fading of adaptive latitudinal clines [56]; the so-called 'Bogert effect' [57]. One thing is to assume that by choosing a specific temperature the organisms mitigate fluctuations in their thermal environment and little selection for temperature-related changes occur as we did, and quite another is to suppose that this behavior helps the survival of the organisms until hereditary variation favored by natural selection in the new environment can be accommodated (Baldwin effect).

To conclude, the demonstration that a standard neo-Darwinian account without learning can easily solve Hinton and Nowlan's [34] harsh task should move "Baldwin boosters" [25] to their winter retreats. Interested colleagues, including ourselves, are invited to come up with a more convincing case for a fascinating and potentially important mechanism.

**Acknowledgments**

We thank Chrisantha Fernando and Stephen R. Proulx for helpful comments on earlier drafts. MS was funded by grant CGL2013-42432 from the Ministerio de Economía y Competitividad (Spain), grant 2014 SGR 1346 from Generalitat de Catalunya, and by the ICREA Acadèmia Program. ES acknowledges financial support from the European Research Council under the European Community's Seventh Framework Programme (FP7/2007–2013/ERC grant agreement no 294332) and the Hungarian National Office for Research and Technology (NAP 2005/KCKHA005). JFF was funded by grant 303979/2013-5 from Conselho Nacional de Desenvolvimento Científico e Tecnológico (CNPq). The funders had no role in study design, data collection and analysis, decision to publish, or preparation of the manuscript.

**Appendix A**

*Epistasis in a single-peaked fitness landscape*

In the scenario without learning the neural connections are also specified by $L$ loci but now with two alleles each, 1 and 0, because in this situation there is no difference between alleles 0 and ?. Since there is no smoothing out of the fitness landscape the fitness function is:

$$w_i = \begin{cases} L & \text{if the organism has all correct alleles 1} \\ 1 & \text{otherwise} \end{cases} \quad (A.1)$$

With $L = 2$ the haplotype frequencies $x_i$ and the corresponding fitness values are given by the following table

|  |  | Locus 1 |  |  |
| --- | --- | --- | --- | --- |
|  |  | 1 | 0 |  |
| Locus 2 | 1 | $x_1 = p_1 p_2 + D$  $w_1 = 2$ | $x_3 = q_1 p_2 - D$  $w_3 = 1$ | $p_2$ |
|  | 0 | $x_2 = p_1 q_2 - D$  $w_2 = 1$ | $x_4 = q_1 q_2 + D$  $w_4 = 1$ | $q_2$ |
|  |  | $p_1$ | $q_1$ | 1 |

$(A.2)$

where $D$ $(D = x_1 - p_1 p_2)$ is the linkage disequilibrium due to the nonrandom association of alleles within haplotypes. The multiplicative fitness epistasis is

$$E' = \ln\left(\frac{w_1 w_4}{w_2 w_3}\right) = 0.6931 \quad (A.3)$$

For the haploid case Felsenstein [58] has shown that directional two-locus selection will tend to generate linkage disequilibrium of the same sign as the multiplicative epistatic parameter. This is illustrated here by iterating the standard recursion equations to calculate gametic frequencies after selection:

$$\begin{aligned} x'_1 &= w_1(x_1 - rD)/\bar{w} \\ x'_2 &= w_2(x_2 + rD)/\bar{w} \\ x'_3 &= w_3(x_3 + rD)/\bar{w} \\ x'_4 &= w_4(x_4 - rD)/\bar{w} \end{aligned} \quad (A.4)$$

where $r$ is the amount of recombination between the two loci and

$$\bar{w} = \sum_{i=1}^{4} w_i x_i. \quad (A.5)$$

Assuming $D_0 = 0$ and $p_1 = p_2 = 0.1$, the linkage disequilibrium over time is plotted in Fig. A1 for two extreme recombination values $(r = 0; r = 0.5)$. $D$ can increase up to near its theoretical maximum 0.25.

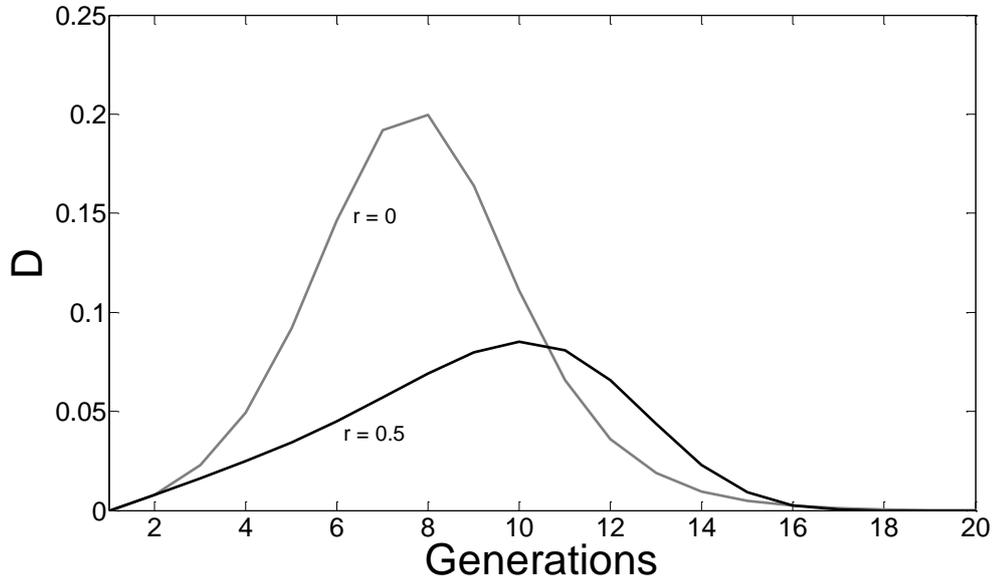

**Fig. A1.** Generation of gametic linkage disequilibrium without learning.

Felsenstein [58] also showed that if the linkage disequilibrium generated by epistatic selection is positive, tighter linkage accelerates the change in allelic frequencies. Therefore, the analytical recursion equations to analyze the evolution of allele frequencies in Hinton and Nowlan's [34] model without learning [37] are expected to grossly overestimate the speed of evolution as they did not take into account the generation of linkage disequilibrium.

**Appendix B**

*Exact recursion equations for the non-learning asexual case*

We will consider the case where learning is absent in order to show the effect of neglecting the generation of linkage disequilibrium due to selection. In this case the genotypes are binary strings of size $L$ because without learning there is no difference between alleles $0$ and $?$. There are $2^L$ different such strings and we denote their frequencies in the infinite population at generation $t$ by $Y_\alpha(t)$, with $\alpha = 1, \ldots, 2^L$. Without loss of generality, we will assume that the correct genotype, i.e. the string $(1, 1, \ldots, 1)$, corresponds to $\alpha = 1$. Recalling that the fitness $w_\alpha$ of all genotypes but the correct one is set to the baseline value $w_{\alpha \neq 1} = 1$, and that the fitness of the correct genotype is set to $w_1 = L$, we can immediately write the recursion equations

$$Y_1(t+1) = \frac{w_1 Y_1(t)}{\sum_\alpha w_\alpha Y_\alpha(t)} = \frac{L Y_1(t)}{1 + (L-1) Y_1(t)}, \tag{B.1}$$

$$Y_{\alpha \neq 1}(t+1) = \frac{w_{\alpha \neq 1} Y_{\alpha \neq 1}(t)}{\sum_\alpha w_\alpha Y_\alpha(t)} = \frac{Y_{\alpha \neq 1}(t)}{1 + (L-1) Y_1(t)}. \tag{B.2}$$

The denominator of the fractions in the right-hand-side of these equations is the average fitness of the population. It is also of interest to calculate the frequency of alleles 1 at generation $t$, which we denote by $p(t)$. To simplify this calculation, let us assume that at generation $t = 0$ the frequencies of all genotypes different from the correct one take on the same value, say $Y_{\alpha \neq 1}(0) = Y_2(0)$. Hence, Eq. (B.2) guarantees that these frequencies will always remain identical, i.e., $Y_{\alpha \neq 1}(t) = Y_2(t)$ for all $t$. With this assumption, it is straightforward to write an equation for $p(t)$

$$p(t) = \frac{1}{L} \sum_\alpha P_\alpha Y_\alpha(t) = Y_1(t) + (2^{L-1} - 1) Y_2(t) = \frac{2^{L-1} [1 + Y_1(t)] - 1}{2^L - 1}, \tag{B.3}$$

were $P_\alpha$ stands for the number of alleles 1 in string $\alpha$ and we have used the normalization condition $Y_1(t) + (2^L - 1) Y_2(t) = 1$. For large $L$, Eq. (B.3) reduces to $p(t) = [1 + Y_1(t)]/2$.

Thus, the study of the case of evolution without learning reduces to finding the solution of Eq. (B.1) that yields the frequency of the correct genotype $Y_1(t)$ as a function of the generation number $t$. For $t \gg 1$ we can rewrite the recursion Eq. (B.1) as the ordinary differential equation

$$\frac{dY_1}{dt} = (L-1) \frac{Y_1(1 - Y_1)}{1 + (L-1) Y_1}, \tag{B.4}$$

the solution of which is simply

$$\frac{Y_1(t)}{[1 - Y_1(t)]^L} = \frac{Y_1(0)}{[1 - Y_1(0)]^L} \exp[(L-1)t]. \tag{B.5}$$

Since we are interested in the regime $t \gg 1$ for which $Y_1(t) \approx 1$, this equation can be further rewritten as

$$Y_1(t) \approx 1 - \frac{1-Y_1(0)}{[Y_1(0)]^L} \exp[-(L-1)t/L]. \tag{B.6}$$

This expression is useful because it allows us to estimate the number of generations needed for $Y_1(t)$ to attain some arbitrary value close to one. In fact, solving Eq. (B.6) for $t$ yields

$$t \approx -\frac{1}{L-1} \ln\left\{ Y_1(0) \left[ \frac{1-Y_1(t)}{1-Y_1(0)} \right]^L \right\}, \tag{B.7}$$

from which we see that $t$ increases with the logarithm of $Y_1(0)$ rather than with a negative power of $Y_1(0)$ as in the case where linkage disequilibrium is neglected [37].

**Appendix C**

*Exact recursion equations for the non-learning asexual case assuming mutation*

We sketch the derivation here solely to emphasize the equivalence between the asexual case and the well-known single-peak fitness landscape of the quasispecies model [59]. The analysis simplifies greatly if one assumes that the relevant feature to distinguish the strings is the number of correct alleles they have without regard to their specific positions in the string. Provided that the initial condition is consistent with this assumption (i.e., all strings with the same number of alleles 1 have the same initial frequencies) the derived recursion equations are exact for the single peak landscape.

We begin by grouping all strings into $L+1$ classes, $i = 0, 1, ..., L$, according to the number of alleles 1 they have. The frequencies of those classes are simply $\Pi_i = \sum_\alpha Y_\alpha \delta_{iP_\alpha}$, where $\delta_{ij}$ is the Kronecker delta. Hence $\Pi_L = Y_1$ since $P_1 = L$ and there is only one string with $L$ alleles 1. The idea is to derive a recursion equation for the class frequencies rather for the string frequencies, which amounts to reducing the number of equations from $2^L$ to $L+1$. The probability that a string with $j$ 1s mutates to a string with $i$ 1s is given by

$$M_{ij} = \sum_{k=a}^{b} \binom{j}{k}\binom{L-j}{i-k}(1-u)^{L-i-j+2k} u^{i+j-2k}, \tag{C.1}$$

where $u$ is the per site probability of mutation. Here, $a = \max(0, i+j-L)$ and $b = \min(i,j)$. In particular,

$$M_{iL} = \binom{L}{i}(1-u)^i u^{L-i}. \tag{C.2}$$

With this in mind we can write the recursion equation for the frequency of strings with $i$ 1s as

$$\Pi_i = \frac{1}{\bar{w}}\left[ LM_{iL}\Pi_L + \sum_{j=0}^{L-1} M_{ij}\Pi_j \right], \tag{C.3}$$

where $\bar{w} = 1 + (L-1)\Pi_L$. Note that this is the equation for Eigen's [59] model in the single-peak fitness landscape. In particular, the reproduction rate of the 'master sequence' (i.e., the all 1s string) is $L$, which is identical to the sequence length. Hence, we expect the error threshold transition to take place at $u_c \approx 1 - 1/L^{1/L} \approx 0.139$ for $L = 20$. Simulation results (not shown) indicate that this figure is a good approximation.

**References**

1. Simpson GG (1953) The Baldwin effect. Evolution 7: 110-117.
2. Baldwin JM (1896) A new factor in evolution. Am Nat 30: 441-451.
3. Morgan CL (1896) On modification and variation. Science 4: 733-740.
4. Osborn HF (1896) Ontogenic and phylogenic variation. Science 4: 786-789.
5. Maynard Smith J (1987) Natural selection: when learning guides evolution. Nature 329: 761-762.
6. Turney P, Whitley D, Anderson RW (1997) Evolution, learning, and instinct: 100 years of the Baldwin effect. Evol Comp 4: iv-viii.
7. Godfrey-Smith P (2003) Between Baldwin skepticism and Baldwin boosterism. In: Weber BH, Depew DJ, editors. Evolution and learning: the Baldwin effect reconsidered. MIT Press, Cambridge. pp 53-67.
8. Griffiths, P. E. 2003. Beyond the Baldwin effect: James Mark Baldwin's "social heredity," epigenetic inheritance, and niche construction. In: Weber BH, Depew DJ, editors. Evolution and learning: the Baldwin effect reconsidered. MIT Press, Cambridge. pp 193-215.
9. West-Eberhard MJ (2003) Developmental plasticity and evolution. Oxford Univ Press, Oxford. 816 p.
10. Crispo E (2007) The Baldwin effect and genetic assimilation: revisiting two mechanisms of evolutionary change mediated by phenotypic plasticity. Evolution 61: 2469-2479.

11. Waddington CH (1953) The"Baldwin effect", "genetic assimilation" and "homeostasis". Evolution 7: 386-387.
12. Huxley JS (1942) Evolution: the modern synthesis. Allen & Unwin, London. 645 p.
13. Mayr E (1963) Animal species and evolution. Harvard Univ Press, Cambridge. 811 p.
14. Dobzhansky T (1970) Genetics of the evolutionary process. Columbia Univ Press, New York. 505 p.
15. Ridley M (2004) Evolution, 3rd edition. Blackwell, Oxford. 784 p.
16. Futuyma DJ (2005) Evolution. Sinauer, Sunderland. 603 p.
17. Barton NH, Briggs DEG, Eisen JA, Goldstein DB, Patel NH (2007) Evolution. CSHL Press, New York. 833 p.
18. Schlichting CD, Pigliucci M (1998) Phenotypic evolution: A reaction norm perspective. Sinauer, Sunderland. 387 p.
19. Avital E, Jablonka E (2001) Animal traditions: behavioural inheritance in evolution. Cambridge Univ Press, Cambridge. 448 p.
20. Schlichting CD, Wund MA (2014) Phenotypic plasticity and epigenetic marking: An assessment of evidence for genetic accommodation. Evolution 68: 656-672.
21. Braendle C, Flatt T (2006) A role for genetic accommodation in evolution? BioEssays 28: 868-873.
22. Laland K, Uller T, Feldman M, Sterelny K, Müller GB, et al. (2014) Does evolutionary theory need a rethink? Nature 514: 161-164.
23. Mitchell M (1996) An introduction to genetic algorithms. MIT Press, Cambridge. 221 p.
24. Back T, Fogel DB, Michalewicz Z editors (1997) Handbook of evolutionary computation. Oxford Univ Press, Oxford. 988 p.
25. Weber BH, Depew DJ editors (2003) Evolution and learning: the Baldwin effect reconsidered. MIT Press, Cambridge. 352 p.
26. Dennett DC (1995) Darwin's dangerous idea. Simon and Schuster, New York. 586 p.
27. Deacon T (1997) The symbolic species: the co-evolution of language and the brain. WW Norton, New York. 528 p.
28. Pinker S (1997) How the mind works. Norton, New York. 672 p.
29. Pinker S, Bloom P (1990) Natural language and natural selection. Behav Brain Sci 13: 707-727.
30. Briscoe EJ (1997) Co-evolution of language and of the language acquisition device. In: Cohen PR, Wahlster W, editors. Proceedings of the Thirty-Fifth Annual Meeting of the Association for Computational Linguistics and Eighth Conference of the European Chapter of the Association for Computational Linguistics. Somerset, New Jersey. pp 418–427.
31. Calvin WH, Bickerton D (2000) Lingua ex machina: reconciling Darwin and Chomsky with the human brain. MIT Press, Cambridge. 312 p.
32. Dor D, Jablonka E (2001) How language changed the genes: toward an explicit account of the evolution of language. In: Trabant J, Ward S, editors. New essays on the origin of language. Trends in linguistics: studies and monographs. Mouton de Gruyter, Berlin. pp 151–175.

33. Yamauchi H (2004) Baldwinian accounts of language evolution. PhD Thesis (Univ Edinburgh, Edinburgh).
34. Hinton G, Nowlan S (1987) How learning can guide evolution. Complex Systems 1: 495-502.
35. Frank SA (2011) Natural selection. II. Developmental variability and evolutionary rate. J Evol Biol 24: 2310-2320.
36. Ancel LW (2000) Undermining the Baldwin expediting effect: does phenotypic plasticity accelerate evolution? Theor Pop Biol 58: 307-319.
37. Fontanari JF, Meir R (1990) The effect of learning on the evolution of asexual populations. Complex Systems 4: 401-414.
38. Wilson DS (1980) The natural selection of populations and communities. Benjamin-Cumings, Menlo Park. 186 p.
39. Alves D, Campos PRA, Silva ATC, Fontanari JF (2000) Group selection models in prebiotic evolution. Phys Rev E 63: 011911.
40. Belew RK (1990) Evolution, learning, and culture: computational metaphors for adaptive algorithms. Complex Systems 4: 11-49.
41. Puentedura RR (2003) The Baldwin effect in the age of computation. In: Weber BH, Depew DJ, editors. Evolution and learning: the Baldwin effect reconsidered. MIT Press, Cambridge. pp 219-234.
42. Sznajder B, Sabelis MW, Egas M (2012) How adaptive learning affects evolution: reviewing theory on the Baldwin effect. Evol Biol 39: 301-310.
43. Fraser A, Burnell D (1970) Computer models in genetics. McGrawHill, New York, 192 p.
44. Comeron JM, Ratnappan R, Bailin S (2012) The many landscapes of recombination in *Drosophila melanogaster*. PLoS Genet 8(10): e1002905.
45. Dennett D (2003) The Baldwin effect: a crane, not a skyhook. In: Weber BH, Depew DJ, editors. Evolution and learning: the Baldwin effect reconsidered. MIT Press, Cambridge. pp 69-79.
46. Dennett D (1991) Consciousness explained. Little, Brown and Company, Boston. 528 p.
47. Orr HA (1999) An evolutionary dead-end? Science 285: 343-344.
48. Maynard Smith J, Szathmáry E (1995) The major transitions in evolution. Oxford Univ Press, Oxford. 360 p.
49. Yeh PJ, Price TD (2004) Adaptive phenotypic plasticity and the successful colonization of a novel environment. Am Nat 164: 531–542.
50. Scheiner SM (2014) The Baldwin effect: neglected and misunderstood. Am Nat 184: ii-iii.
51. Dawson A (2008) Control of the annual cycle in birds: endocrine constraints and plasticity in response to ecological variability. Phil Transact Roy Soc B 363: 1621-1633.
52. Maynard Smith J (1965) Obituary. Prof. J. B. S. Haldane, F.R.S. Nature 206: 239-240.
53. Duckworth RA ( 09) The role of behavior in evolution: a search for mechanism. Evol Ecol 23: 513-531.
54. Corning PA (2014) Evolution 'on purpose': how behaviour has shaped the evolutionary process. Biol J Linn Soc 112: 242-260.

**55.** Mayr E (1960) The emergence of evolutionary novelties. In Tax S, editor. Evolution after Darwin. Univ Chicago Press, Chicago. pp 349-380.
**56.** Castañeda LE, Balanyà J, Rezende EL, Santos M (2013) Vanishing chromosomal inversion clines in *Drosophila subobscura* from Chile: is behavioral thermoregulation to blame? Am Nat 182: 249-259.
**57.** Huey RB, Hertz PE, Sinervo B (2003) Behavioral drive versus behavioral inertia in evolution: a null model approach. Am Nat 161: 357–366.
**58.** Felsenstein J (1965) The effect of linkage on directional selection. Genetics 52: 349-63.
**59.** Eigen M (1971) Self organization of matter and the evolution of biological macromolecules. Naturwissenschaften 10:465-523.